\documentclass[epj]{svjour}
\usepackage{graphics}
\usepackage{graphicx}
\usepackage{indentfirst}
\usepackage{amsmath,amssymb,array,cases,graphicx,algorithmic}
\usepackage{,booktabs}
\usepackage{tabularx}
\usepackage{cite}

\begin{document}

\title{Symmetric coexisting attractors in a novel memristors-based Chua’s chaotic system}
\author{Shaohui. Yan\thanks{\emph{mortal\_sysh@163.com} }, Zhenlong. Song, Wanlin. Shi, Weilong. Zhao,}
%
%
%
\institute{College of Physics and Electronic Engineering, Northwest Normal University, Lanzhou, Gansu 730070,China}
\date{Received: date / Revised version: date}
%
\abstract{
In this paper, based on the classic Chua’s circuit, a charge-controlled memristor is introduced to design a novel four-dimensional chaotic system. The complex dynamics of the novel chaotic system such as equilibrium points, stability, dissipation, bifurcation diagrams, Lyapunov exponent spectra and phase portraits are investigated. By varying the initial conditions of the system, it is found from numerical simulations that the system shows some dynamics of great interests including double-wings chaotic attractors, coexisting periodic-chaotic bubbles, asymmetric and symmetric coexisting attrators. The results show that the novel circuit system has extreme multistablity.
\PACS{
      { }{ Chua’s Chaotic system， Memristor， Bifurcation analysis， Coexisting attractors， Symmetry}
     } 
} 
\maketitle
\section{Introduction}
\label{intro}
Nonlinear electronic circuits provide an effective way to produce chaotic behavior \cite{[1]}\cite{[2]}\cite{[3]}. Chua’s circuit is a simple nonlinear chaotic circuit made by Professor Cai Shaotang in1983 \cite{[4]}\cite{[5]}\cite{[6]}\cite{[7]}. Chua's circuit contains four basic elements and a nonlinear resistance, but there have been hundreds of research papers. The details of Chua's circuit have been deeply investigated including topology, numerical simulation, dynamical characterizations and physical phenomena \cite{[8]}\cite{[9]}\cite{[10]}\cite{[11]}\cite{[12]}. Because of the Chua's circuit system has the characteristics of extreme initial value sensitivity and good pseudo-randomness, which has been widely used in science and engineering, \cite{[13]}, robotics \cite{[14]}, random generator implementation \cite{[15]}, secure communication and even image encryption \cite{[16]}, and synchronous encryption \cite{[17]}. The coexistence of multiple attractors have been found in many of nonlinear systems and electronic circuits\cite{[18]}\cite{[19]}\cite{[20]}\cite{[21]}. In general, the appearance of coexisting attractors is associated with system’s symmetry and depends closely on system initial conditions. Chaotic system with multiple attractors is able to deliver more complexity in chaos-based engineering applications such as neural networks \cite{[22]}, image encryption \cite{[23]}, control system \cite{[24]} and random number generator \cite{[25]}. Therefore, chaotic system with coexisting attractors has become a considerable interest at present. In 1971, according to the completeness principle of circuit theory Chua predicted the fourth electronic component and named memristor, which has the unique property of remembering the past electric charge \cite{[26]}\cite{[27]}. The memristor was created by Hewlett Packard laboratory, whose resistance was characterized by the nonlinear constitutive relation between charge and flux \cite{[28]}. Because of nature non-linearity, plasticity of memristors, simple circuit topology and complex dynamical behaviors, a lot of attraction have been attention to memristor-based applications \cite{[29]}\cite{[30]}, especially neural networks \cite{[31]}, write / read circuits \cite{[32]}, image encryption \cite{[33]}, voice encryption \cite{[34]}. From the above discussion, these studies mainly focus on the using of flux-controlled memeristor. In 2008, HP laboratory announced the charge-controlled memristor, which is shown more practical application than the flux-controlled memristor. However,the charge-controlled memristors have been little reported in the literature. Thus, chaotic system with the charge-controlled memristors has important research value in the dynamics with the variation of parameters and the multistability phenomenon depending on different initial conditions. This paper is organized as follows: In Section 2. the mathematical model of the Chua's deformation circuit is completed. In Section 3, the complex dynamical behaviors are numerically revealed by means of bifurcation diagrams, Lyapunov exponents, phase portraits and symmetric coexisting attractors’ behaviors. By changing the initial value, the Lyapunov exponents actually show the symmetry of zero point. Finally, the conclusion is given in Section 4.
\section{The memristor-based chaotic model}
\label{sec2:0}
In this section, a memristor-based chaotic circuit system is introduced and its mathematical model is discussed.
\subsection{Circuit description}
\label{sec2:1}
\begin{figure}
\begin{center}
  \includegraphics[scale=0.5]{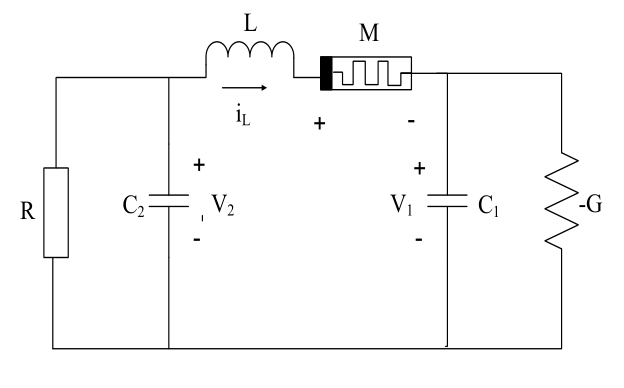}
  \caption{Circuit schematic of the floating ground type memristive chaotic circuit}
\label{fig:1}
\end{center}
\end{figure}

A floating ground type memristor chaotic circuit is shown in fig.\ref{fig:1}. Its specific structure design is used a nonlinear memristor and an inductor in series between two capacitors, forming a connection method that allows the memristor to float the ground type \cite{[35]}. The memristive circuit also contains one linear resistor R and one negative conductance minus G. The memristor model employs charge-controlled piece wise linear (PWL) model, the nonlinear relationship between the variable $q$ is shown (\ref{test}). The corresponding memristor value is defined as (\ref{test2}). Where $q$ is the charge, $M(q)$ are the charge-controlled resistance value, n is the resistance of memristor with $q$ greater than 1 and m is the resistance of memristor with $q$ less than or equal to 1.
\begin{equation} \label{test}
f(q)=nq+0.5(m-n)(|q+1|-|q-1|)
\end{equation}


\begin{equation}\label{test2}
M(q)= \frac{df(q)}{dq}
\begin{cases}
m & |q|\leqslant 1\\
n & |q|>1
\end {cases}
\end{equation}
\subsection{Mathematical model}
\label{sec2:2}
\begin{figure}
\begin{center}
  \includegraphics[scale=0.4]{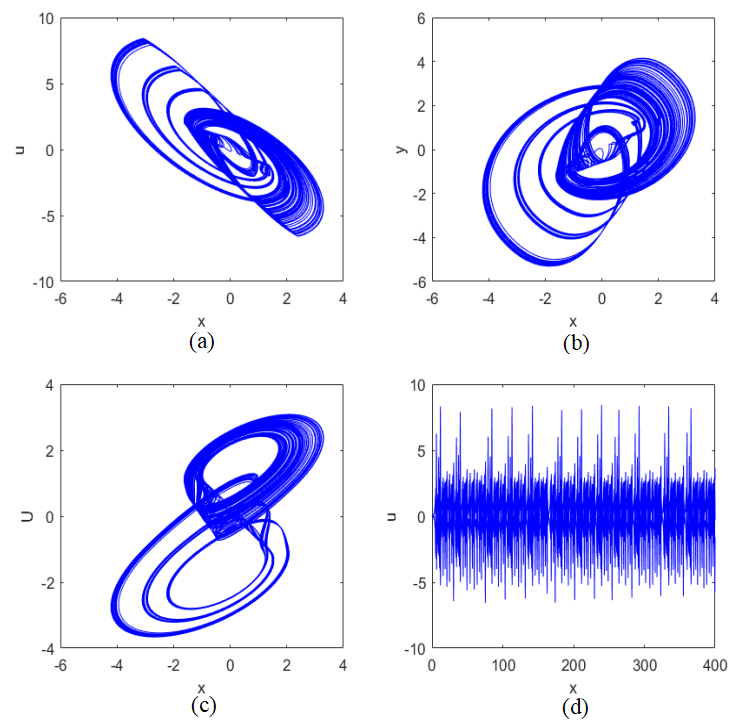}
  \caption{Phase portraits in (a) $x-z$ plane, (b) $x-y$ plane, (c) $x-u$ plane, (d) the time-domain waveform of state variable $x$, with the parameters $\alpha = 9, \beta = 3, \gamma = 4.6, \delta = 15, e = 0.9, m = 0.8, n = -1.5$ and the initial conditions $x(0) = 0.001, y(0) = 0.02, z(0) = 0, u(0) = 0.1.$}
\label{fig:2}
\end{center}
\end{figure}

A novel chaotic circuit based on the proposed memristor is constructed as showed  fig.\ref{fig:1}. By applying Kirchhoff’s laws, the circuit nonlinear differential equations can be derived as follows
\begin{equation}\label{test3}
 \left\{
\begin{aligned}
\frac{dv_1}{dt} & =  \frac{1}{c_1}(i_L+G_Nv_1) \\
\frac{dv_2}{dt} & =  \frac{1}{c_2}(-v_2-i_L) \\
\frac{di_L}{dt} & =  \frac{1}{L}(v_2-v_1-i_1M(q)) \\
\frac{dq}{dt} & =  i_l \\
\end{aligned}
\right.
\end{equation}
By normalizing the state variables and circuit parameters are set as
\begin{equation} \label{test4}
x=v_1,y=v_2,z=i_L,u=q,\alpha=\frac{1}{C_1},\beta=\frac{G_N}{C_1},e=\frac{1}{L}
\end{equation}
among them
\begin{equation}\label{test5}
M(u)=
\begin{cases}
m & |q|\leqslant 1\\
n & |q|>1
\end {cases}
\end{equation}
The introduction of dimensionless variable ( x, y, z, u ) yield the following state equation
\begin{equation}\label{test6}
 \left\{
\begin{aligned}
\dot{x} & =  \alpha z +\beta x\\
\dot{y} & =  -\gamma y-\delta z \\
\dot{z} & =  e (y-x-M(u)z) \\
\dot{u} & =  z\\
\end{aligned}
\right.
\end{equation}
It implies that (\ref{test6}) is four-dimensional memristive chaotic system and has four state variables. The system parameters are chosen as follows, $ \alpha=9, \beta=3, \gamma=4.6, \delta=15, e=0.9, m=0.8, n=-1.5$ and the initial condition (x(0), y(0), z(0), u(0)) = (0.001, 0.02, 0, 0.1)). The largest Lyapunov exponents is calculated LE = 0.72. The phase portraits and time-domain waveform show typical dynamic behaviors of the system (\ref{test6}), as show in  fig.\ref{fig:2}. including asymmetry of the chaotic attractor. The typical system parameters to simulate the fourth order dynamic behavior are summarized in table.\ref{tab:1}.
\section{Dynamical behaviors of the system}
\label{sec3:0}
\subsection{Equilibrium point and its stability }
\label{sec3:1}
The equilibrium point of the mathematical model (\ref{test6}) can be expressed as $p=(\overline{x},\overline{y},\overline{z},\overline{u})$whose values are solved by the following equations:
\begin{equation}\label{test7}
 \left\{
\begin{aligned}
0 & =  \alpha z +\beta x\\
0 & =  -\gamma y-\delta z \\
0 & =  e (y-x-M(u)z) \\
0 & =  z\\
\end{aligned}
\right.
\end{equation}
Apparently, the equilibrium point can be gotten by setting $A=\{(x, y, z, u)| x=y=z=0,u=\gamma\}$, where $'\gamma'$ is a constant. It means that this memristive chaotic system has a line of equilibrium corresponding to the u-axis. The Jacobian matrix in the equilibrium point can be easily given as:
\begin{equation}\label{test8}
P(\lambda)=det \hspace{3pt} t(1\lambda-J)=\lambda^3+A\lambda^2+B\lambda+C
\end{equation}
Where
\begin{equation}\label{test9}
A=\gamma+M(u)e-\beta, \hspace{3pt} B=2\alpha e+\gamma M(u)e, \hspace{3pt} C=e(\gamma M(u)\beta+\alpha \beta-\gamma \alpha)%
\end{equation}
According to the characteristic equation and known parameters $\alpha=9, \beta=3, \gamma= 4.6, \delta=15, e=1, m=0.8, n=-1.5$, when$|u|\leqslant 1 $ the corresponding eigenvalues of the system are:
\begin{equation}\label{test10}
\lambda_1=0, \hspace{3pt} \lambda_2=0.5777, \hspace{3pt} \lambda_3,\lambda_4=-1.4139\pm4.7827i%
\end{equation}
The root of $\lambda_2$ is positive real constant, whereas the $\lambda_3$ and $\lambda_4$ are a pair of imaginary roots, which indicate that (\ref{test10}) is an unstable saddle focus with an index of 1. Because they have two complex conjugate roots with negative real part and one positive root \cite{[36]}. When $|u|>1$ the corresponding eigenvalues of those system is:
\begin{equation}\label{test11}
\lambda_1=0, \hspace{3pt} \lambda_2=-1.2271, \hspace{3pt} \lambda_3,\lambda_4=0.9385\pm4.4746i%
\end{equation}
Also, the eigenvalues is an unstable saddle focus with an index of 2. The saddle focus equilibrium of index 2 is the premise for the generation of vortex motion, while the saddle focus equilibrium of index 1 is the basis for the formation of bond bands between the coils \cite{[37]}. Therefore, it can be concluded that the equilibrium point is always unstable and conforms to the conditions for the generation of chaos \cite{[38]}.
\subsection{Dissipative analysis }
\label{sec3:2}
\begin{figure}
\begin{center}
  \includegraphics[scale=0.3]{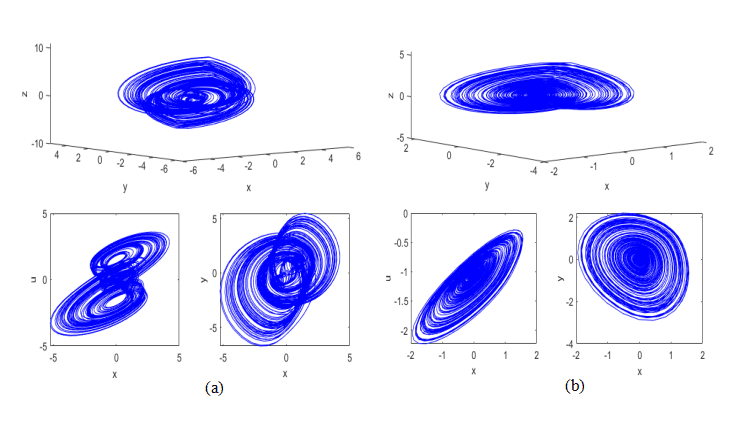}
  \caption{Two and three dimensional phase portraits with different , (a) $e = 1.5$, Row 1 is the phase portrait in the $x-y-z$ plane; Row 2, column 1 is the phase portrait in the $x-u$ plane; Row 2, column 2 is the phase portrait in the $x-y$ plane. (b) $e =1.5$, Row 1 is the phase portrait in the $x-y-z$ plane; Row 2, column 1 is the phase portrait in the $x-u$ plane; Row 2, column 2 is the phase portrait in the $x-y$ plane. The parameters are chosen as $\alpha = 9; \beta = 3; \gamma = 4.6, \delta = 15, m = 0.8, n= -1.5$, and the initial conditions are $x(0) = 0.001, y(0) = 0.02, z(0) = 0, u(0) = 0.1.$}
\label{fig:3}
\end{center}
\end{figure}

For the memristive chaotic system (\ref{test6}), its vector field divergence is:
\begin{equation}\label{test12}
\nabla V=\frac{d\dot{x}}{dx}+\frac{d\dot{y}}{dy}+\frac{d\dot{z}}{dz}+\frac{d\dot{u}}{du}=\beta-\gamma-M(u)e
\end{equation}
Fixing $\alpha = 9, \beta = 3, \gamma = 4.6, \delta = 15, m = 0.8, n = -1.5 and e = 1.5$, the $\nabla V$ is obviously negative, when setting the $|u|\leqslant 1$ , as shown in (\ref{test13}).
\begin{equation}\label{test13}
\nabla V=
\begin{cases}
-2.8 & |u|\leqslant 1\\
0.65 & |u|>1
\end {cases}
\end{equation}
\begin{equation}\label{test14}
\nabla V=
\begin{cases}
-1.6-0.8e & |u|\leqslant 1\\
-1.6+1.5e & |u|>1
\end {cases}
\end{equation}
When $e$ is a variable parameter, it can obtain a divergence relation (\ref{test14}). Based on the analysis above, divergence relation of the system is specified for table.\ref{tab:2}. The system is dissipative when $|u|\leq 1$ and $e>-2$, which is also dissipative when $|u|>1$ and $e< 16/15$. Ideally, the phase space orbit after many folds and stretches can eventually be confined to a limited subset and eventually fixed in an attractive domain forming an attractor. Then another type of transition changes in the number of attractor’s wings occur in this system. As $e$ goes on, the phase portrait transfers from double-wings becoming single-wing attractors, as shown in  fig.\ref{fig:3}.
\subsection{Lyapunov exponents spectrum and bifurcation analysis}
\label{sec3:3}
\begin{figure}
\begin{center}
  \includegraphics[scale=0.4]{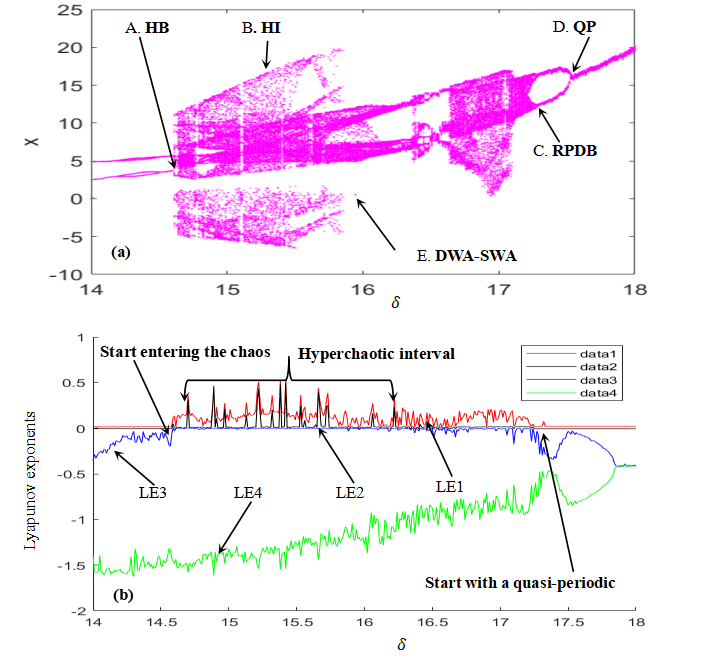}
  \caption{The diagram of (a) bifurcation and (b) Lyapunov exponents with $ 14 ≤ \delta ≤ 18 $. The parameters are chosen as $\alpha = 9, \beta = 3, \gamma = 4.6, e = 0.9, m = 0.8, n = -1.5$ and the initial conditions $x(0) = 0.001, y(0) = 0.02, z(0) = 0, u(0) = 0.1.$}
\label{fig:4}
\end{center}
\end{figure}

In this subsection, the dynamic characteristics of system (\ref{test6}) are presented through bifurcation diagram, Lyapunov exponents spectrum and phase diagrams. The coexisting attractors mean multiple attractors with their own domains of attraction with respect to different initial variable. The numerical calculations are performed by Adomian decomposition method (ADM) with a fixed step size is 0.01s \cite{[39]}. For $\alpha = 9, \beta = 3, \gamma = 4.6, e = 0.9, m = 0.8, n = -1.5$, the bifurcation and Lyapunov exponents diagram of system (\ref{test6}) with respect to parameter $\delta\in\{14, 8\}$ starting from the initial value (x(0), y(0), z(0), u(0)) = (0.001, 0.02, 0, 0.1)) are shown in  fig.\ref{fig:4}, which shows the system is a weak hyperchaotic behaviors. In the hyperchaotic discontinuity region, the maximum Lyapunov index LE1 is always greater than zero, while LE2 keeps jumping between greater than zero and less than zero in  fig.\ref{fig:4}(b). When $δ > 17.42$, the dynamic behavior of the system (\ref{test6}) shifts from chaos to periodic behavior and then to the limit cycle with the evolution of $\delta$. When $\delta = 17.33$, it can be seen from the bifurcation diagram in  fig.\ref{fig:4}(a) that the system enters reverse periodic doubling bifurcation, which is well confirmed on the Lyapunov exponents spectrum. When $\delta\in\{14, 14.8\}$ (except some narrow periodic windows), the maximum Lyapunov exponent is not positive and the system is periodic orbit. When $\delta\in\{14.6, 17.42\}$, the maximum Lyapunov exponent is positive, so the system is chaotic. When $\delta\in\{17.48, 18\}$, the maximum Lyapunov exponent is approximately zero, which means the system returns periodic orbit.\par
\begin{figure}
\begin{center}
  \includegraphics[scale=0.3]{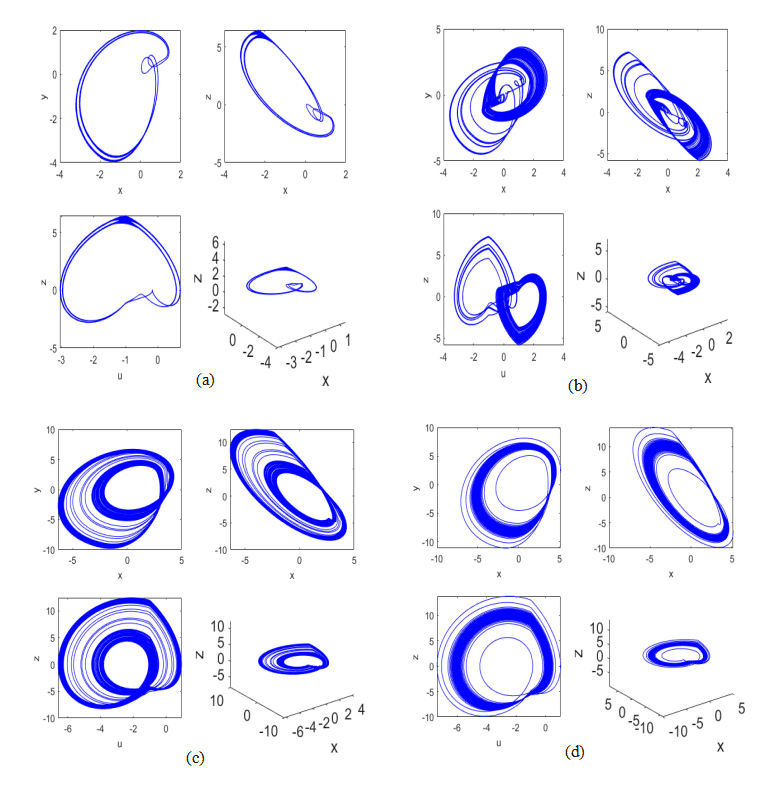}
  \caption{Two and three dimensional phase portraits with different $\delta$. (a) $\delta =14.5$, Row 1, column 1 is the phase portrait in the $x-y$ plane; Row 1 column 2 is the phase portrait in the $x-z$ plane; Row 2, column 1 is the phase portrait in the $u-z$ plane; Row 2, column 2 is the phase portrait in the $x-y-z$ plane. (b) $\delta = 14.8$, Row 1, column 1 is the phase portrait in the $x-y$ plane; Row 1, column 2 is the phase portrait in the $x-z$ plane; Row 2, column 1 is the phase portrait in the $u-z$ plane; Row 2, column 2 is the
phase portrait in the $x-y-z$ plane. (c) $\delta = 17.33$, Row 1, column 1 is the phase portrait in the $x-y$ plane; Row 1, column 2 is the phase portrait in the $x-z$ plane; Row 2, column 1 is the phase portrait in the $u-z$ plane; Row 2, column 2 is the phase portrait in the $x-y-z$ plane. (d) $\delta = 17.6$, Row 1, column 1 is the phase portrait in the $x-y$ plane; Row 1, column 2 is the phase portrait in the $x-z$ plane; Row 2, column 1 is the phase portrait in the $u-z$ plane, Row 2, column 2 is the phase portrait in the $x-y-z$ plane. The parameters are chosen as $\alpha=9, \beta=3, \gamma=4.6, \delta=15, m=0.8, n=-1.5$ and the initial conditions are $x(0) = 0.001, y(0) = 0.02, z(0) = 0, u(0) = 0.1$}
\label{fig:5}
\end{center}
\end{figure}

：
As shown in  fig.\ref{fig:4}(a), the points of A, B, C and D correspond to the phase portraits of  fig.\ref{fig:5}.(a),  fig.\ref{fig:5}(b), fig.\ref{fig:5}(c) and  fig.\ref{fig:5}(d), respectively. A is Hop-bifurcation (HB), B is hyperchaos interval(HI), C is reversed period doubling bifurcated point (RPDB), D is starting point of quasi-periodic (QP) and E is the point where the double-wings attractor becomes single-wing attractor (DWA-SWA). For the fixed parameter of $\delta$, the phase portraits demonstrating different dynamical behaviors are provided, as shown in  fig.\ref{fig:5}.  fig.\ref{fig:5}(a) illustrates single-wing periodic state at $\delta = 14.5$.  fig.\ref{fig:5}(b) illustrates double-wings asymmetric attractors at $\delta = 14.8$.  fig.\ref{fig:5}(c) illustrates quasi-periodic state at $\delta = 17.33$, which shows that the window period exists.  fig.\ref{fig:5}(d) illustrates limit cycle at $\delta = 17.65$. The results in  fig.\ref{fig:5}. are consistent with those in fig.\ref{fig:4} The system is extreme multistability, namely, the coexistence of attractors for different system parameter δ is distinctly observed.\par
\begin{figure}
\begin{center}
  \includegraphics[scale=0.4]{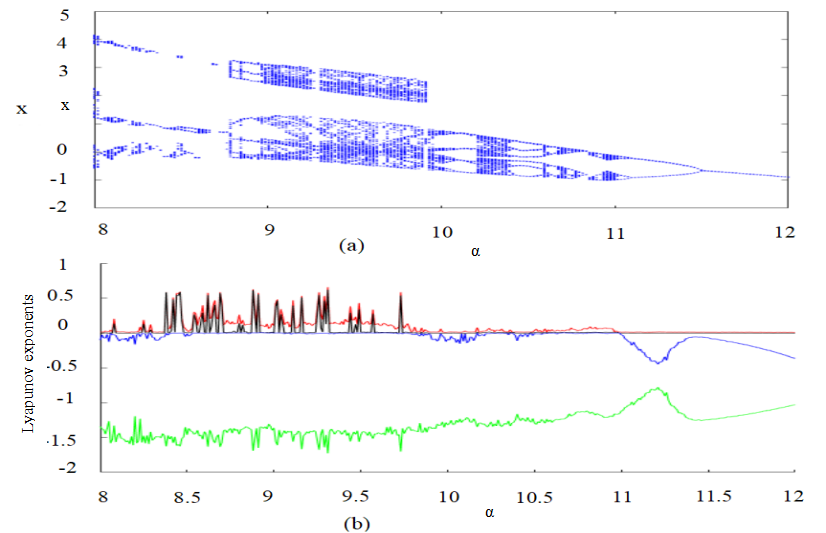}
  \caption{Corresponding bifurcation diagram (a) and Lyapunov exponents spectrum (b) with respect to parameter $ \alpha $. Fix $ \beta = 3, \gamma = 4.6, m = 0.8, n = -1.5, \delta = 15 $ and $ e = 0.9 $, the initial values are $ x(0) = 0.001, y(0) = 0.02, z(0) = 0, u(0) = 0.1.$
}
\label{fig:6}
\end{center}
\end{figure}

\begin{figure}
\begin{center}
  \includegraphics[scale=0.55]{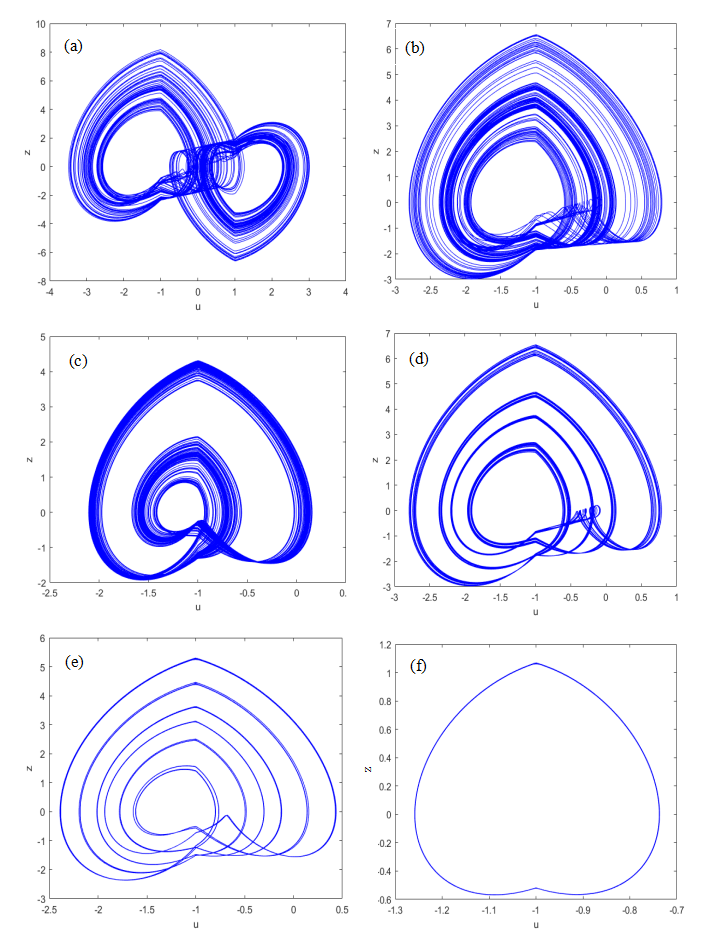}
  \caption{ phase portraits obtained for different values of $c$ in: (a) hyperchaotic attractor at $c = 9.19$, (b) chaotic attractor at $c = 10.22$, (c) chaotic attractor at $c = 10.7$, (d) quasi periodic at $c = 9.968$, (e) quasi periodic at $c = 10.41$, (f) Limit cycle with period-1 at $c = 11.7$. The parameters are chosen as $\alpha=9, \beta=3, \gamma=4.6,\delta=15, m=0.8, n=-1.5$ and initial conditions are $x(0) = 0.001, y(0) = 0.02, z(0) = 0, u(0) = 0.1.$}
\label{fig:7}
\end{center}
\end{figure}

：
To investigate the effect of parameter α on the dynamical behaviors of system (6), we let $\beta = 3, \gamma =  4.6, m = 0.8, n = -1.5, \delta = 15, e = 0.91$ and $\alpha\in\{8, 12\}$. Here, the initial values are chosen as X0 (0.001, 0.02, 0, 0.1) and the corresponding bifurcation diagram and Lyapunov exponents spectrum are shown in fig.\ref{fig:6}(a) and fig.\ref{fig:6}(b), respectively. Clearly, the maximum Lyapunov exponent is positive and the system can generate chaos for $\alpha \in \{9, 9.8\}\cup \{10.2, 10.4\}\cup \{10.6, 11\}$. For the phase diagram of fig.\ref{fig:7}(a), fig.\ref{fig:7}(b) and fig.\ref{fig:7}(c), there apparently emerge some periodic windows sandwiched in the chaotic band. While the system is periodic for $\alpha\in\{9.9, 10.2\}\cup\{10.4, 10.6\}\cup\{11, 12\}$ ( as shown in fig.\ref{fig:7}(d), fig.\ref{fig:7}(e) and fig.\ref{fig:7}(f), respectively).
\subsection{Symmetrical rotation of coexisting attractors}
\label{sec3:4}
There is a absorbing phenomenon in this chaotic system that has hardly been proposed in other memristive chaotic systems. It has shown that single-wing attractor and double-wings attractor appear to symmetrical rotation in this paper. The mentioned single-wing and double-wings attractor are aimed at the $x-y$ phase diagrams. That is a single-wing attractor in one position can be symmetrical rotated to generate single-wing attractor in different locations. Usually, varying the parameters of system may induce the phenomenon of rotation. However, the rotation in this system is caused by different memristor initial conditions, which is really different from the previous literatures. Therefore, this phenomenon can be called the rotation of coexisting attractors.\par
\begin{figure}
\begin{center}
  \includegraphics[scale=0.35]{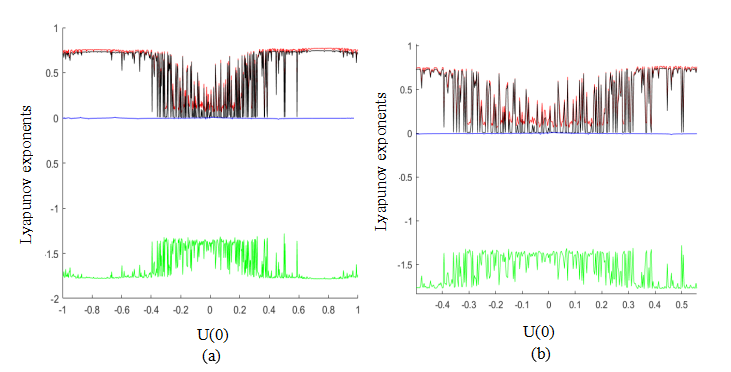}
  \caption{symmetry Lyapunov exponent diagram. (a) Lyapunov exponent spectra, the parameter $u(0)$=-1 to 1. (b) Lyapunov exponent spectra, the parameter $u(0)$=-0.5 to 0.55. Fix $\beta = 3, \gamma = 4.6, m = 0.8, n = 1.5, \delta = 15 and e = 0.9$, the initial values are $x(0) = 0.001, y(0) = 0.02, z(0) = 0, u(0)$.}
\label{fig:8}
\end{center}
\end{figure}

：
Setting system parameters as $\alpha = 9, \beta = 3, \gamma = 4.6, e = 0.9, m = 0.8, n = -1.5, \delta = 15$ and the initial condition of three state variables as $x(0) = 0.001, y(0) = 0.2$ and $z(0) = 0$, then the memristor initial condition $u(0)$ is taken as the Lyapunov exponent spectra parameter, as shown in fig.\ref{fig:8}(a). It can be acquired that the system undergoes a complex alternation of numerous chaotic and periodic states with $u(0)$ increasing from -1 to 1. For the sake of clarity, the middle complex part of the Lyapunov exponent spectra diagram is magnified, as exhibited in fig.\ref{fig:8}(b).\par
\begin{figure}
\begin{center}
  \includegraphics[scale=0.4]{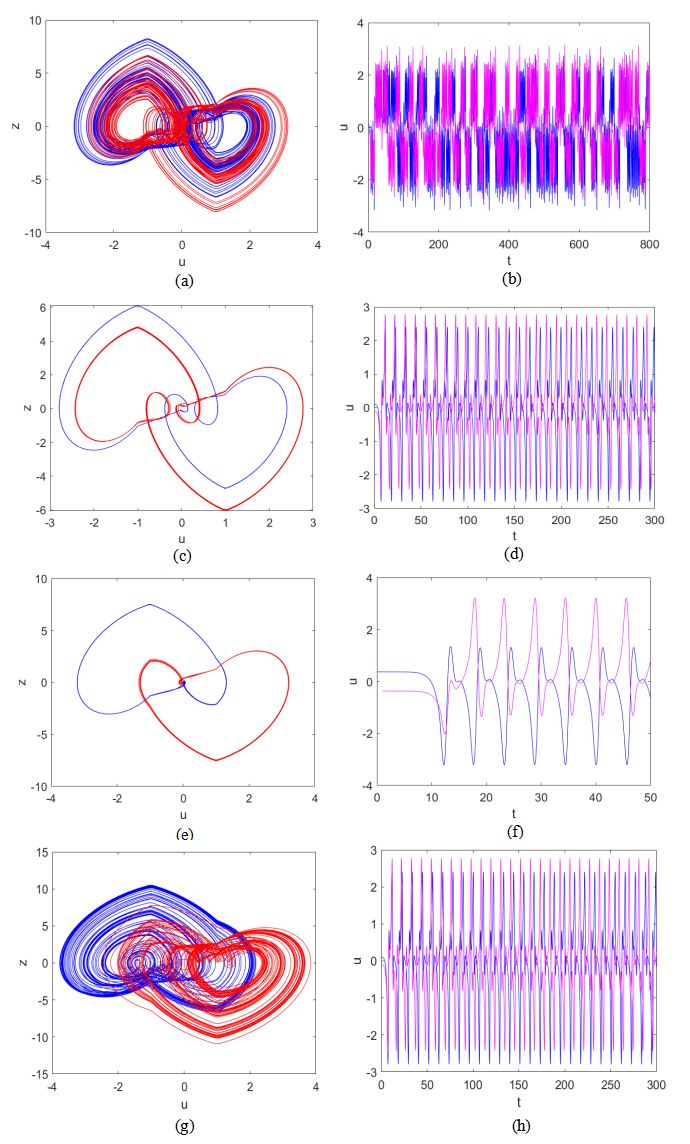}
  \caption{Phase plane orbits in u-z plane and the corresponding time series of system (6) with parameters $\alpha = 9, \beta = 3, \gamma = 4.6, m = 0.8, n = -1.5, \delta = 14, X0 (0.001, 0.02, 0 , u)$ and $X1 (0.001, 0.02, 0, -u)$. (a) coexisting chaotic attrators and (b) time series of $u $ at $e = 1, u = 0.1$, (c) coexisting periodic-2 attractors (d) time series of u at e = 0.9, u = 0.1, (e) coexisting periodic-1 and (f) time series  of $ u $ at $ e = 0.9. u = 0.37 $, (g) quasi periodic at and (h) time series of $ u $ at $ e = 1, u = 0.37 $.}
\label{fig:9}
\end{center}
\end{figure}

：
Obviously, system (\ref{test6}) is symmetric in $u-z$ axis. Consequently, a pair of symmetric attractors can be generated by altering the initial conditions. Let $\alpha = 9, \beta = 3, \gamma = 4.6, m = 0.8, n = -1.5, \delta = 14$ and initial values $X0 (0.001, 0.02, 0, u)$ and $X1 (0.001, 0.02, 0, -u)$. The system coexists two chaotic attractors, when parameters set $e = 1$ and $u = 0.1$ respect to initial value X0 (blue) and X1 (red) as shown in fig.\ref{fig:9}(a) and fig.\ref{fig:9}(b). If $e = 0.9$ and $u = 0.1$, the system coexists one periodic-2 attractors with respect to initial value X0 (blue) and X1 (red) as shown in fig.\ref{fig:9}(c) and fig.\ref{fig:9}(d). If e = 0.9 and u = 0.37, the system coexists one periodic-1 attractors and one chaotic attractors with respect to initial value X0 (blue) and X1 (red) as shown in fig.\ref{fig:9}(e) and fig.\ref{fig:9}(f). If $e = 1, u=0.37$, the system is quasi periodic with respect to initial value X0 (blue) and X1 (red) as shown in fig.\ref{fig:9}(g) and fig.\ref{fig:9}(h). And the periodic attractors place on symmetrical position of the phase space with similar structure. As described above, it shows that the novel chaotic system exists coexisting attractors and manifests the presence of multistability.\par：
\section{Conclusion}
\label{sec4:0}
In this paper, we proposed the charge-controlled memristor of chaotic circuit system. The dynamical behaviors of the novel system are highly complex and sensitive with regard to different circuit parameters and initial conditions. Through using nonlinear analysis methods, it was found that some complex dynamical phenomena including state transition, chaos generation and degradation, and asymmetrical and symmetrical coexisting attractors. Simultaneously, we presented reasonable explanations for these dynamical behaviors. Hence, the novel chaotic system has great prospects in engineering applications such as image encryption, secure communication and neural networks.

%
%

%
%

\newpage
\newpage

\begin{table}[t]
\caption{\label{tab:1}
The fixing control parameters for simulations.}
\begin{center}
\begin{tabular}{clccc}
\toprule  
Parameters & Values \\
\hline
  $\alpha$ & 9 \\
  $\beta$ & 3 \\
  $\gamma$ & 15\\
   $e$ & 0.9 \\
  $m$ & 0.8 \\
  $n$ & -1.5\\
\toprule  
\end{tabular}
\end{center}
\end{table}

\begin{table}[t]
\caption{\label{tab:2}
Divergence relation of system (6).}
\begin{center}
\begin{tabular}{clccc}
\toprule  
Parameters u & Parameters e & divergence & type \\
\hline
 $|u| \leq 1$ & $e > -2$          & $\bigtriangledown V<0$ & Dissipation type \\
\\
 $|u| \leq 1$ & $e < -2$          & $\bigtriangledown V>0$ & -\\
\\
 $|u| > 1$    & $e > \frac{16}{15}$ & $\bigtriangledown V>0$ & - \\
\\
 $|u| > 1$    & $e < \frac{16}{15}$ & $\bigtriangledown V<0$ & Dissipation type \\
\toprule  
\end{tabular}
\end{center}
\end{table}
\end{document}